\documentclass[aps,prb,twocolumn,superscriptaddress,amsmath,showpacs,floatfix,longbibliography]{revtex4-1}

\usepackage{graphicx} 
\usepackage{color}

\definecolor{darkgreen}{rgb}{0,0.4,0}

\usepackage{color}
\usepackage{amsfonts,amssymb,amsmath}
\usepackage{tabularx,hhline}
\usepackage{tikz}

\def\DTO{Dy$_{2}$Ti$_{2}$O$_{7}$}
\def\YTO{Y$_{2}$Ti$_{2}$O$_{7}$}
\def\HTO{Ho$_{2}$Ti$_{2}$O$_{7}$}
\def\LuTO{Lu$_{2}$Ti$_{2}$O$_{7}$}
\def\TbTO{Tb$_{2}$Ti$_{2}$O$_{7}$}
\def\calH{\mathcal{H}}

\newcommand{\bn}{$\boxed{\varnothing}$}
\newcommand{\bbp}{$\boxed{+}$}
\newcommand{\bbm}{$\boxed{-}$}
\newcommand{\bpp}{$\boxed{\begin{array}{c}+\\+\end{array}}$}
\newcommand{\bmm}{$\boxed{\begin{array}{c}-\\-\end{array}}$}

\begin{document}

\title{ Emergent Electrochemistry in Spin Ice: Debye--H\"{u}ckel Theory and Beyond}

\author{V. Kaiser}
\affiliation{Universit\'e de Lyon, ENS de Lyon, Universit\'e Claude Bernard, CNRS, 
Laboratoire de Physique, F-69342 Lyon, France} 
\affiliation{Max-Planck-Institut f\"ur Physik komplexer Systeme, 01187 Dresden, Germany}
\affiliation{Center for Systems Biology Dresden, Max Planck Institute of Molecular Cell Biology and Genetics, 01307 Dresden, Germany}

\author{J. Bloxsom}
\affiliation{London Centre for Nanotechnology and Department of Physics and Astronomy, University College London,
London WC1H 0AH, United Kingdom}

\author{L. Bovo}
\affiliation{London Centre for Nanotechnology and Department of Physics and Astronomy, University College London,
London WC1H 0AH, United Kingdom}
\affiliation{Department of Innovation and Enterprise, University College London, 90 Tottenham Court Road, London W1T 4TJ, United Kingdom.}

\author{S. T. Bramwell}
\affiliation{London Centre for Nanotechnology and Department of Physics and Astronomy, University College London,
London WC1H 0AH, United Kingdom}

\author{P. C. W. Holdsworth}
\affiliation{Universit\'e de Lyon, ENS de Lyon, Universit\'e Claude Bernard, CNRS, 
Laboratoire de Physique, F-69342 Lyon, France} 

\author{R. Moessner}
\affiliation{Max-Planck-Institut f\"ur Physik komplexer Systeme, 01187 Dresden, Germany}

\date{\today}

\begin{abstract}
The low-temperature picture of dipolar spin ice in terms of the Coulomb fluid of its fractionalised magnetic monopole excitations has allowed analytic and conceptual progress far beyond its original microscopic spin description. 
Here we develop its thermodynamic treatment as a `magnetolyte', a fluid of singly and doubly charged monopoles, an  
analogue of the electrochemical system  ${\rm 2 H_2O = H_3O^+ +OH^-  = H_4O^{2+} + O^{2-}}$, but with  perfect symmetry between oppositely charged ions. 
For this lattice magnetolyte, we present an analysis based on Debye--H\"uckel theory, which is accurate at all temperatures 
and incorporates `Dirac strings' imposed by the microscopic ice rule constraints 
at the level of Pauling's approximation. 
Our results are in close agreement with the specific heat from numerical 
simulations as well as new experimental measurements with an 
improved lattice correction, which we present here, on the spin ice materials~\HTO~and~\DTO. 
Our study of the magnetolyte  shows how electrochemistry can emerge in non-electrical systems.
We also provide new experimental tests of Debye--H\"uckel theory and its extensions. 
The application of 
our results also yields insights into the electrochemical behaviour of water ice and liquid water, which are closely
related to the spin ice magnetolyte. 
\end{abstract}


\maketitle

\section{\label{sec: intro}
Introduction
        }
        
Capturing charge correlations in a Coulomb fluid is a notoriously challenging problem. 
Long range interactions mean that the equilibrium state is only stabilised through the build up of charge screening correlations,
so that approaches beyond simple perturbation theory are required, even at the highest temperatures and lowest charge concentrations. Debye and H\"uckel's \cite{Debye1923} approximate solution of the problem, along with Bjerrum's extension \cite{Bjerrum1926} to include association, established a controlled theory that remains the cornerstone of theoretical approaches\cite{Mayer1950,Justice1977,Ebeling1982}. Comparatively recently, Fisher and Levin \cite{Fisher1993,Levin1996} extended the theory to cover the whole temperature-density phase diagram of a model fluid, while Kobelev {\it et. al.} \cite{Kobelev2002} treated lattice systems. 
Practically though, making contact with experiments in electrolytes, over a wide range of charge concentration, requires a more elaborate description, including the coupling between electrostatics and physico-chemical effects. 
For example, Pitzer's model \cite{Pitzer1973A,Pitzer1973B} is based on Debye--H\"uckel theory, but relies on several fitting parameters to include both solvation and steric effects.

\begin{figure*} [ht]
\centering{\includegraphics[scale=1]{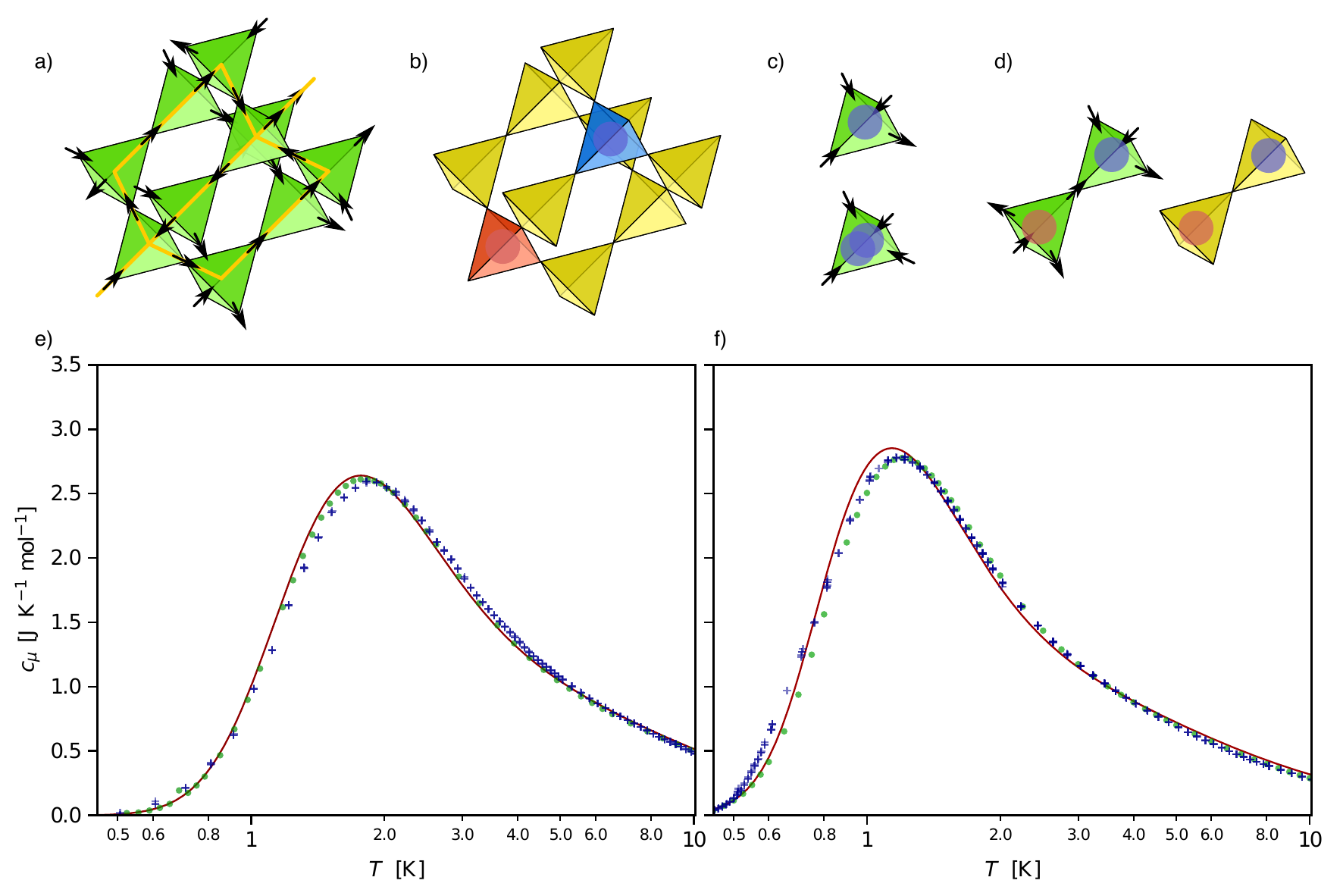}}
\caption[$\;\;$ Specific heat]{{\textbf{Specific heat -- experiment}: }
Monopole model of spin ice and theory versus experiment. (a) The spin ice state on the pyrochlore lattice forms a vacuum for magnetic monopoles in the magnetolyte model. (b) Neglecting the string network yields a diamond lattice electrolyte. (c) Magnetic monopoles of  single charge (upper) and double charge (lower). (d)  Charge pairs in the magnetolyte and electrolyte. (e,f) The experimental specific heat (blue crosses) of \HTO (e) and \DTO (f) 
as compared with simulations (green circles) and 
and Debye--H\"uckel theory with monopole pairing included  (red line).}
\label{intro-figure}
\end{figure*}

Spin ice has provided an unexpected setting for the study of Coulomb physics.
The low temperature state of spin ice with its associated Pauling entropy \cite{Harris1997,Bramwell2001,Ramirez99} provides an effective ground state from which pointlike defects are thermally excited \cite{Ryzhkin2005,Castelnovo2008}. Using the dumbbell model \cite{Castelnovo2008} of dipolar spin ice \cite{denHertog2000}, these fractionalised particles \cite{Fulde2002} were shown to interact via a magnetic Coulomb potential, giving a `magnetolyte' of magnetic monopoles \cite{Castelnovo2008}. 

In this paper we show that the spin ice magnetolyte broadens the scope of Coulomb systems, 
providing a model Coulomb fluid on a lattice whose thermodynamic properties are accessible to experiment, 
not only at low temperature, but over a broad range of thermodynamic variables, 
with only few undetermined or phenomenological parameters involved, 
it spans both the high and low temperature limits in particular. 

Spin ice is particularly attractive in this context as it naturally yields a Coulomb fluid in the grand canonical ensemble in which the external parameter is the chemical potential for monopole creation, rather than charge or fluid density. It is perfectly symmetric,
due to the time reversal symmetry of magnetism, which eliminates ion-specific effects, making it relatively easy to model. Further, as the charges are quasi particles confined to a solid state environment, pressure and volume are effectively decoupled from the Coulomb thermodynamics while the underlying lattice structure greatly facilitates entropy calculations.  
As a consequence we are able to adapt Debye--H\"uckel and association theory to the magnetolyte, allowing detailed comparison of theory both with simulation  and with experiments on spin ice materials.  

Debye--H\"uckel theory for spin ice was first formulated in Ref.~[\onlinecite{Castelnovo2008}]. Here we develop a full thermodynamic description of the dumbbell model that allows a complete quantitative comparison of theory, experiment and simulation. This goes beyond the previous work~[\onlinecite{Castelnovo2008}] in three important ways.  
First, see Fig.~1$(a)$, we take into account the fact that  the monopole vacuum state is actually an ensemble of configurations of close packed and constrained magnetic moments with finite entropy density \cite{Harris1997,Bramwell2001,Ramirez99} that we include into the Coulomb fluid at the Pauling level of approximation \cite{Pauling1935}. This incorporates the fragmentation of the the magnetic moments into a monopole and a vacuum contribution for arbitrary monopole concentrations \cite{Brooks2014}. By suppressing this contribution one recovers a simple lattice electrolyte (Fig.~1$(b)$) that is studied for comparison. Second (Fig.~1$(c)$), we
allow for not only singly charged~[\onlinecite{Castelnovo2008}], but also doubly charged monopoles, in analogy with an electrochemical system of the form:  
\begin{equation}\label{scheme}
{\rm 2 H_2O \rightleftharpoons H_3O^+ +OH^- \rightleftharpoons H_4O^{2+} + O^{2-}}\ .
\end{equation}
Considering double charges permits us to access the full temperature range (see App.~\ref{AppTheories}).
Third (Fig.~1 $(d)$), as well as formulating Debye--H\"uckel theory, we refine it through the systematic inclusion of neutral bound charge pairs for both the lattice electrolyte and the magnetolyte. This refinement is in the spirit of Bjerrum's theory, but is specifically adapted to the spin ice magnetolyte. The associated lattice electrolyte is treated using the technique of Ref.~[\onlinecite{Kobelev2002}].  

The net result of these developments is an approximate, yet highly accurate, analytic solution of a three dimensional spin model with long range interactions -- the dipolar spin ice model \cite{denHertog2000} (see below). Our solution approaches an exact description of dipolar spin ice over a restricted parameter range, where the Debye--H\"uckel linear approximation is valid. The parameters of the canonical spin ice materials Ho$_2$Ti$_2$O$_7$ (HTO) and Dy$_2$Ti$_2$O$_7$ (DTO) lie within this range at most temperatures; hence, as shown in Figs.~$1(e)$ and $(f)$, our theory describes their experimental and simulated specific heat with only very small systematic corrections (see App.~\ref{AppErrors}).  It should be noted that the experimental data of Figs.~1 $(e)$ and $(f)$  improves on some previous measurements, in that the non-magnetic background component has been estimated with precision up to high temperature, allowing the close confrontation of experiment, theory and simulation.  

The striking agreement between experiment, simulation and theory shown in Figs. 1 $(e)$ and $(f)$ confirms the existence of an emergent electrochemistry in spin ice over a full range of charge concentrations. To our knowledge a quantitative demonstration of the applicability of Debye--H\"uckel theory to specific heat measurements has not previously been achieved. 

The famous limiting law for the the activity coefficient,  which was discovered experimentally~\cite{Lewis1921}, before it was derived theoretically by Debye and H\"uckel~\cite{Debye1923} (see eqn. (\ref{limit-law}))  
is implicit in the data. However, going beyond this limit has proved difficult as
a detailed comparison of theory and experiment for electrolytes has generally been hampered by imprecise knowledge of parameters, as well as the difficulty of accounting for strong correlations. In the case of the spin ice magnetolyte, the emergent nature and perfect symmetry of the magnetic charge largely eliminates these problems.

The models studied and developed in this paper, and tested numerically, are broadly relevant to a number of experimental systems in electrochemistry and magnetism. The lattice electrolyte model could essentially describe a weak (solid or liquid) electrolyte in which the dissociating ions are not strongly correlated with the `solvent'. The magnetolyte model could, through the analogy between electro- and magnetostatics, describe water ice or, approximately, water itself.  In these cases the solvent is correlated with the ions through Dirac strings or, equivalently, hydrogen--bonded chains. One result of our work is to show how such correlations may be factored into the Pauling entropy, to allow for a standard electrochemical description at the level of thermodynamics.   

The remainder of the paper justifies our main result, Fig.~1 $(e)$ and $(f)$, and exposes a number of other notable details. It is organised as follows. In the next section we develop the model magnetolyte, highlighting its specific characteristics compared to the lattice electrolyte. In
section 3 we present electrolyte and magnetolyte thermodynamics, developing  equations of state within the Debye--H\"{u}ckel approximation. The limits of the theory are given and extensions to it at low temperature, using a pair approximation, are discussed in App.~\ref{AppPairing}. In section 4 we test the theory by comparing specific data from both simulations and our new experiments. Conclusions are drawn in section 5.

\section{Models: from electrolyte to magnetolyte}

Spin ice~\cite{Harris1997,Bramwell2001} is a corner sharing network of tetrahedra forming a pyrochlore lattice of localized, Ising like magnetic moments, as shown in  Fig.~\ref{intro-figure} $(a)$.  
The magnetic degrees of freedom transform, to an excellent approximation, into a fluid of magnetic monopoles \cite{Castelnovo2008} at the centres of the tetrahedra, forming a diamond lattice. The monopole charges emerge from flux lines, `Dirac strings', that connect the sites and constrain charge movement. The transformation is made by extending the spins into infinitesimally thin needles carrying net neutral dumbbells of charge that touch at the diamond lattice sites \cite{Moller2006,Castelnovo2008,Jaubert2011a,Castelnovo2018}. The total charge, $Q_\alpha$ accumulated at the site 
 $\alpha=1\ldots N_0$, is the sum of the four dumbbell charges arriving at tetrahedron $\alpha$. Allowed values are thus $Q_\alpha=0,\pm Q,\pm2Q$, where $Q$ is the monopole charge, so that the Hamiltonian reads
\begin{equation}
\calH-\calH_0={1\over{2}}\sum_{\alpha \ne \beta}{\mu_0 Q_\alpha Q_\beta \over{4\pi r_{\alpha\beta}}} -\mu N^{i} - \mu_2 N^{i}_2 \ ,
\label{dumbbell}
\end{equation} 
with $r_{\alpha\beta}$ the distance separating sites $\alpha$ and $\beta$, and  
$\mu_0$ the magnetic permeability. The chemical potentials, $\mu$ and $\mu_2$ are for single and double monopoles,
 $N^{i}$ ($N_2^{i}$) are the respective number of (double) monopoles/charges for a given 
 microstate, and $\calH_0$ is the  energy of the charge vacuum \cite{Castelnovo2008}. 
Strictly speaking, $\mu$ and $\mu_2$ are excess terms \cite{Frenkel2002}, 
defined with respect to a reference state at $\mu=\mu_2=0$, 
which has random spin configurations constraining dense but globally charge neutral charge configurations. See the Conclusions section of this work for further discussion.
 
The mapping to a grand canonical fluid means that the independent thermodynamic variables 
are $T$, $\mu$ and $\mu_2$, which together with the diamond lattice constant $a$ and the monopole charge \cite{Castelnovo2008}
completely specify the problem. In spin ice materials $\mu_2=4\mu$, but interesting physics can appear for other choices of this ratio \cite{Brooks2014}. Setting $\mu_2=\infty$ imposes $N^i_2=0$, which we define as the primitive model.
 
Neglecting the string network leads to a standard lattice electrolyte \cite{Castelnovo2011}, while taking into account its additional features defines the magnetolyte system (which is equivalent to a description of water ice without D/L defects). In the absence of any charge, the Dirac string network carries a finite `Pauling' entropy \cite{Harris1997,Pauling1935}, corresponding to the ensemble of spin configurations satisfying the ice rules with two spins pointing in and two out of each tetrahedron.
It is accurately accounted for by the Pauling approximation, $S_P\approx k_BN_0\ln\left(\frac{3}{2}\right)$ (Ref.~[\onlinecite{Pauling1935}]) providing a monopole vacuum with finite entropy. Configurations with a finite monopole concentration generally maintain some of this entropy, associated with the free space between the quasi-particles, so that each charge state should be supplemented  with an entropic  
weight given by the number of spin microstates consistent with it. Certain microstates of the lattice electrolyte are forbidden by this procedure. For example two adjacent $Q=2$  charges cannot be nearest neighbours in the magnetolyte. These are high energy states and so not important in the fluid phase, but they can have consequences for monopole crystallization at high density \cite{Borzi2014}. In the following we are able to add this entropic weight at the mean field level, which does not take such correlations into consideration.

A generic phase diagram for lattice electrolytes on bi-partite structures allowing non-frustrated ionic crystals has been studied in detail by Kobelev{\it et al.}~\cite{Kobelev2002}, albeit for fields confined to lattice edges. For the primitive model, a fluid phase gives way to a crystalline phase via a transition that is either first or second order, separated by a tri-critical point \cite{Dickman1999}. The magnetolyte with both single and double charges has a similar phase diagram, with crystallization to a double monopole zinc-blende structure \cite{Brooks2014,Borzi2013,Borzi2014}. The monopole-monopole interactions driving this evolution across the phase diagram can be parameterised by the interaction ratio, $\zeta=\frac{u_a}{|\mu|}$, where $u_a=\frac{\mu_0 Q^2}{4\pi a}$ is the Coulomb energy scale for a nearest neighbour pair of charges. Monopole crystallization occurs for  $\zeta=\frac{2}{\alpha}=1.22$, where $\alpha$ is the Madelung constant for a diamond lattice~\cite{Brooks2014}, providing an upper bound for the stability of the fluid phase. For spin ice materials DTO and HTO,  $\zeta=0.71$ and $\zeta=0.54$ respectively, placing them away from the phase boundary \cite{Melko2004} yet far from the non-interacting limit.

The vacuum entropy implies that the idealised dumbbell model violates the third law of thermodynamics in that the entropy remains finite at the absolute zero of temperature. Experiments on both water ice\cite{Giauque1936,Salzmann2011} and spin ice \cite{Ramirez99} show a corresponding residual entropy. The spin systems of real spin ice materials are considered to be metastable \cite{Pomaranski2013} below some low temperature (estimated to be at least $0.3$ K for spin ice), but in the temperature range considered in this paper they accurately approximate the dumbbell model at equilibrium, as our results confirm.

 \section{\label{sec: Free}
Coulomb fluid Thermodynamics}

\subsection{Grand Potential}

The electrolyte and magnetolyte free energies are of the form
\begin{equation}\label{Free-E}
\Omega = U_C  -\mu N-\mu_2 N_2 - ST,
\end{equation}
where $U_C$, $N$ and $N_2$ are thermally averaged values for the Coulomb energy, the number of monopoles and of double monopoles respectively.

Following Ryzhkin~\cite{Ryzhkin2005}, one can write an approximate expression for the vertex entropy of the magnetolyte by considering each type of vertex as a species of indistinguishable objects. 
\begin{equation}
W=\left(\frac{1}{2}\right)^{2N_0} \frac{N_0!}{N_1! N_2! \dots N_{16}!} 
, 
\end{equation}
where $N_a$ is the number of vertices of type $a = 1,\ldots,16$. The prefactor $\left( 1/2 \right)^{2N_0}$ takes into account the compatibility of the spins shared between neighboring vertices. 

Each vertex configuration corresponds to charge $0,\pm Q, \pm 2Q$, with six 2in-2out spin ice configurations corresponding to charge zero, four each of 3in-1out (and four 3out-1in) configurations corresponding to charge $Q$ ($-Q$) and one all-in (all-out) vertex corresponding to charge $2Q$ ($-2Q$). For a system of $N=nN_0$ monopoles and $N_2=n_2N_0$ double monopoles in a system of $N_0$ sites one can hence set 
 $N_1=N_2=\dots N_6=(1-n-n_2)N_0/6$, $N_7=N_8=\dots N_{14}=nN_0/8$ and $N_{15}=N_{16}=n_{2}N_0/2$. It follows that 
\begin{eqnarray}
S = -k_B N_0 
 {\bigg\{}&n& \ln \left( \frac{n}{2} \right) +n_2 \ln \left( {2n_2} \right) \\ \nonumber
	&+& 
	(1-n-n_2) \ln  \left(1-n-n_2\right)  \\ \nonumber
	&+& (1-n-n_2) \ln \left( \frac{2}{3}  \right){\bigg\}}.
	\label{entropy}
\end{eqnarray}
This formula elegantly separates the entropy into a monopole term and a vacuum term: the last term approximates to the vacuum entropy and setting $n=n_2=0$ yields the Pauling entropy, $S_p=k_B N_0 \ln\left( 3/2 \right)$. The first three terms correspond to the entropy of a  lattice gas with both single and double charges. Setting $n_2=0$ and excluding the vacuum entropy gives the primitive electrolyte entropy
\begin{equation}
S_e=-k_BN_0 \left[n \ln (n/2) +(1-n) \ln (1-n)\right],
\label{electro}
\end{equation}
used in reference [\onlinecite{Castelnovo2011}]. For the double monopoles, the vertex weights modify their contribution compared to a free lattice gas, giving a contribution of $n_2\ln (2n_2)$ rather than the $n_2\ln(n_2/2)$ one might have expected. A final check can be made at high temperature, where one expects the single and double monopole concentrations to approach $n_\infty=\frac{1}{2}$ and $n_{2{(\infty)}}=\frac{1}{8}$ respectively. Plugging in these numbers yields the full entropy of $2N_0$ uncorrelated Ising degrees of freedom, $S_\infty=2N_0k_B \ln 2$. 

Armed with this expression for the entropy we can find equations of state for the monopole fluid, $n(\mu,\mu_2,T)$, $n_2(\mu,\mu_2,T)$ by minimising  eqn. (\ref{Free-E}) with respect to $n$ and $n_2$:
\begin{eqnarray}\label{rho-SI}
n&=&\frac{{\frac{4}{3}}\exp(\beta\tilde{\mu})}{ 1+ {\frac{1}{3}}[4\exp(\beta\tilde{\mu})+ \exp(\beta\tilde{\mu}_2)]}, \\ \nonumber
n_2&=&\frac{{\frac{1}{3}}\exp(\beta\tilde{\mu}_2)}{ 1+ {\frac{1}{3}}[4\exp(\beta\tilde{\mu})+ \exp(\beta\tilde{\mu}_2)]},
\end{eqnarray}
where
\begin{eqnarray}
\tilde{\mu}=\mu-k_BT\ln( \gamma),  \; \tilde{\mu}_2=\mu_2-k_BT\ln (\gamma_2),
\end{eqnarray}
and where $\gamma, \; \gamma_2$ are the activity coefficients of the fluid:
\begin{equation}\label{gamma-def}
k_BT\ln (\gamma_i)=\frac{1}{N_0}\frac{\partial U_C}{\partial n_i}.
\end{equation}
One can see that the interactions reduce the energy scale for the inclusion of monopoles at finite density: $\vert \tilde{\mu} \vert < \vert \mu \vert$ and $\gamma<1$, leading to an increased monopole concentration compared to the non-interacting gas in the ratio $1/\gamma\;$~\cite{Kaiser2013}. 

From this, all thermodynamic quantities for the monopole fluid can be calculated. For example, the magnetic specific heat transforms, in this representation to
\begin{eqnarray}\label{cmu_calculation}
C_{\mu,\mu_2}&=&\left(\frac{\partial}{\partial T}\right)(U_C-\mu N -\mu_2 N_2 ),\\ \nonumber
&=&-N_0\left[ \tilde{\mu}\left(\frac{\partial n}{\partial T}\right) + \tilde{\mu}_2\left(\frac{\partial n_2}{\partial T}\right)\right].
\end{eqnarray}
Hence, if one can deal successfully with the Coulomb energy, one can give a complete self-contained description of the magnetolyte fluid in which the spin and magnetic charge degrees of freedom have been included independently, rather in the spirit of the gauge mean field theories used to study quantum spin liquids~\cite{Savary2012}. 

The lowest order approximation is to neglect the Coulomb interaction altogether, giving a non-interacting lattice fluid apart from hard core repulsions.
In this case $\tilde{\mu}$ and $\tilde{\mu}_2$ are equal to the respective chemical potentials and the problem is trivially solved.
This is equivalent to a single tetrahedron approximation for the NNSI model~\cite{Kaiser2014} with $J_{\rm eff}=-\mu/2$.
The specific heat of the NNSI is accurately described by the single tetrahedron model (although not the susceptibility \cite{Jaubert2013}) everywhere in the spin ice phase.
For $\mu>0$, the non-interacting monopoles crystallise via an order by disorder transition to the all-in-all-out phase~\cite{Borzi2014}, but this transition is not captured using the Pauling approximation for the entropy, eqn. (\ref{entropy}).
In the next section we go beyond the non-interacting case, adapting  Debye-H\"{u}ckel theory (see e.g. Refs.~\onlinecite{Levin2002,Kaiser2014}) to the magnetolyte.

\subsection{\label{sec: DH theory}
Debye--H\"{u}ckel theory}
           
Debye--H\"{u}ckel theory~\cite{Debye1923,Levin2002,Kaiser2014} uses the linearized Poisson-Boltzmann equation to go beyond mean field theory,  predicting a correlation induced electrostatic potential, $\psi(r)$, a distance $r$ from a test charge $q=zQ$, with $z$ an arbitrary constant:
\begin{eqnarray}
\psi(r\ge a) &=&  z\left(\frac{\mu_0 Q}{4\pi r}\right)
\frac{\exp\left[-(r-a)/\ell_D\right]}{1+a/\ell_D} \\ \nonumber
 \ell_D&=&\sqrt{\frac{k_BT}{Q^2\rho_I\mu_0}},
\label{eq:DH Bjerrum}
\end{eqnarray}
where $\ell_D$ is the Debye length. The short distance cut off is, in our case, the lattice spacing $a$ of the diamond lattice, $\rho=N/V$, $\rho_2=N_2/V$ are the volume densities of charges and $\rho_I=\rho+4\rho_2$ is the interaction strength \cite{Levin2002} for the magnetolyte with single and double charged monopoles. The test charge induces a charge cloud in its vicinity of opposite sign, whose extension is controlled by $\ell_D$. The Coulomb energy is the energy required to place the test charge in the  induced potential. It can be calculated using the Debye charging procedure in which the charge on each site is built up adiabatically for fixed particle correlations. Setting $Q(\lambda)=\lambda Q$, the Coulomb energy for the test charge at infinitesimally small $\lambda$ is defined 
\begin{eqnarray}
\delta u(\lambda)&=& -z\lambda Q \psi(a,\lambda)\\ \nonumber
 &=& - z^2\left(\frac{\mu_0Q^2}{4\pi a}\right)\frac{\lambda^2}{1+(a\lambda/\ell_D)}.
\end{eqnarray}
 This expression can now be integrated from $\lambda=0$ to $\lambda=1$ to find the Coulomb energy of the test particle, $z^2 u^{DH}$. Taking the test charge to be a single ($z=1$) or a double ($z=2$) monopole gives the internal energy $U_C^{\rm DH}=N_0 n_Iu^{DH}$, with $n_I=n+4n_2$ being the ionic strength:
\begin{eqnarray}
U_C^{\rm DH}=
 - \frac{N_0 k_BT}{6\pi\sqrt{3}}\bigg[\ln\left(1+\frac{a}{\ell_D}\right)
 &-&\left(\frac{a}{\ell_D}\right)\\ \nonumber
 &+&\frac{1}{2}\left(\frac{a}{\ell_D}\right)^2\bigg].
\end{eqnarray}
To convert the extensive variable from volume to $N_0$ we have used the volume per diamond lattice site, $\tilde{v}=8a^3/3\sqrt{3}$~\cite{Castelnovo2011}. Note that, as  $\ell_D \propto 1/\sqrt{n}$,  $U_C^{\rm DH} \sim n^{3/2}$ at low monopole density, contrary to the $n^2$ behaviour typical of mean field descriptions of short range systems. 

Minimizing $\Omega$ with respect to $n$ and $n_2$ gives the effective chemical potentials 
\begin{eqnarray}
\tilde{\mu} &=&\mu+\Delta^{\rm DH}, \; \tilde{\mu}_2=\mu_2 + 4\Delta^{\rm DH},\\ \nonumber
\Delta^{DH}&=&k_BT \frac{\ell_T}{\ell_D+a},
\end{eqnarray}
and activity coefficients, 
\begin{equation}\label{limit-law}
\gamma=\exp {(-\beta\Delta^{\rm DH})} ,\; \gamma_2=\exp {(-4\beta\Delta^{\rm DH})}\, .
\end{equation}
Here, it is convenient to introduce the Bjerrum length, $\ell_T = \frac{\mu_0 Q^2}{8\pi k_BT}$, at which the Coulomb interaction per monopole is equal to the thermal energy scale. The limiting law~\cite{Lewis1921,Debye1923} for low ionic strength $(1-\gamma) \propto \mathrm{const.}\sqrt{n_I}$, follows from eqn. (\ref{limit-law}).

Putting $\tilde{\mu}$ and $\tilde{\mu}_2$ into Eq.~\eqref{rho-SI} and solving  self-consistently for the densities~\cite{Castelnovo2011} gives the  Debye--H\"{u}ckel equations of state $n(\mu,\mu_2,T)$ and $n_2(\mu,\mu_2,T)$ from which all thermodynamic quantities follow.

The details of the calculation can be considerably simplified by setting $\mu_2=-\infty$ and restricting to the primitive (14-vertex)  magnetolyte which can be compared in detail with the primitive electrolyte of reference \cite{Castelnovo2011}. This is justified for spin ice materials at 
low temperatures where the double monopoles can be neglected and is a practical simplification over the whole temperature range. 

\subsection{Limits of validity \& charge pairing}

Restricting to the primitive model for simplicity, the Poisson-Boltzmann equation for the induced potential $\psi(r)$ is:
\begin{equation}
\nabla^2 \psi(\vec r)= -\frac{\mu_0  \rho Q}{2} \left[\exp{(-\beta Q \psi(\vec r))}- \exp{(\beta Q \psi(\vec r))}\right],
\end{equation}
where the equilibrium charge (volume) density for each species, $\rho^{\pm}=\pm \frac{Q\rho}{2}$. This is solved in Debye--H\"{u}ckel theory by keeping the linear terms in the exponential. This is a very poor approximation for $r<\ell_T$ so that the theory essentially ignores excess charge at distances less than this\cite{Guggenheim1959}.
The validity of the Debye--H\"{u}ckel $U_C$ as the leading contribution to the Coulomb energy therefore depends on the contribution made by such near neighbours and the ratio of lengths $\frac{\ell_T}{\ell_D}$ is a good small parameter for this.
At high temperature $\ell_T \rightarrow 0$ while $\ell_D$  diverges as $a\sqrt{\frac{k_BT}{u_a}}$. At low temperature one can use a pair approximation and treat near neighbour pairs as a species in chemical equilibrium. Their contribution to internal energy scales as $n^2$ at small density, compared to $n^{3/2}$ for the Debye--H\"{u}ckel contribution\cite{Bjerrum1926,Justice1977,Levin1996,Ebeling2002}. 

As a consequence one expects the theory to be valid at both low and at high temperatures. 
The Debye--H\"{u}ckel contribution to thermodynamic observables is measured by the activity coefficient, or $\Delta^{DH}$, for which the low and high temperature limits are:
\begin{equation}
\Delta^{DH}(T\rightarrow 0) \sim u_a\sqrt{\frac{u_a n}{k_BT}},\; \Delta^{DH}(T\rightarrow \infty) \sim u_a.
\end{equation}
As the low temperature limiting law gives $\Delta^{DH}$ varying as $\sqrt{n}$ only, one finds significant and experimentally observable contributions even for small charge concentrations~\cite{Lewis1921}. At high temperature $\Delta^{DH}$ is temperature independent, yet finite, illustrating the importance of screening even in this limit.

The short range ionic pairing neglected by Debye--H\"{u}ckel theory generates a contribution to the activity coefficient linear in ionic strength \cite{Ebeling2002}.
We develop pairing approximations for both the electrolyte and magnetolyte, whose details we give in App.~\ref{AppPairing}, and whose results appear in Fig.~\ref{intro-figure} $(e)$ and $(f)$.
For spin ice, our method enumerates the partition sum of two neighbouring tetrahedra (corresponding to 7 spins, i.e. $2^7=128$ vertex states), with the statistical weights adjusted by Bjerrum-like association constants. 

This approach is specific to spin ice, because its short-range structure differs from that of a lattice electrolyte. 
The specificity of pairing contrasts with and underscores the universality of Debye--H\"{u}ckel theory.
Moreover, the improved match between theory and experiment (Fig.~\ref{intro-figure}), achieved by adding the pairing correction, reveals that the small remaining discrepancy between DH theory and the experimental spin ice specific heat is also due to electrostatic correlations, with only a small part accounted for by the error of the Pauling approximation. This confirms spin ice as an experimental realisation of a symmetric lattice electrolyte.

\section{Tests of the theory}

\subsection{Magnetolyte simulations}

\begin{figure} [ht]
\centering{\includegraphics[scale=1.0]{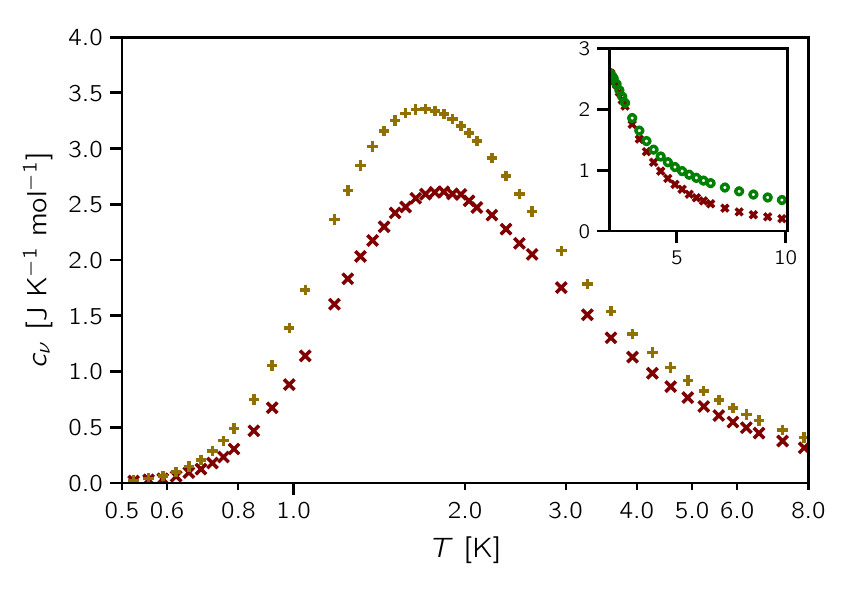}}
\caption[$\;\;$ Specific heat]{{\textbf{Specific heat}: simulation data {\it vs} $T$ for the primitive electrolyte (yellow crosses) and primitive magnetolyte (red crosses) for $\zeta=0.54$, as for HTO. Inset: primitive and full (16-vertex) magnetolyte (green circles) illustrating the effect of double monopoles on specific heat above 2 K.
}}
\label{fig:C_challenge}
\end{figure}

In Fig.~\ref{fig:C_challenge} we show specific heat data for systems with $\mu=-5.7$ K and $\zeta=0.54$, as for HTO. We show simulation data for both the primitive electrolyte and magnetolyte (see App.~\ref{AppSimulations} for methods). Notice that the areas under the curves are significantly different. This is  a consequence of the constrained magnetolyte having a significantly different entropy change going from low to high temperature. The total entropy at high and low temperatures, which can be estimated from eqn. (\ref{entropy}), are $S_\infty/N_0k_B= \ln{7/2}$ and $S_0/N_0k_B=\ln{3/2}$ for the magnetolyte and $S_\infty/N_0k_B= \ln{3}$ and $S_0/N_0k_B=0$ for the electrolyte, respectively. The inset shows the effect of including double monopoles to the magnetolyte. As can be seen, they modify the specific heat from $2$ K and above. 

\begin{figure*} [ht]
\centering{\includegraphics[scale=1.0]{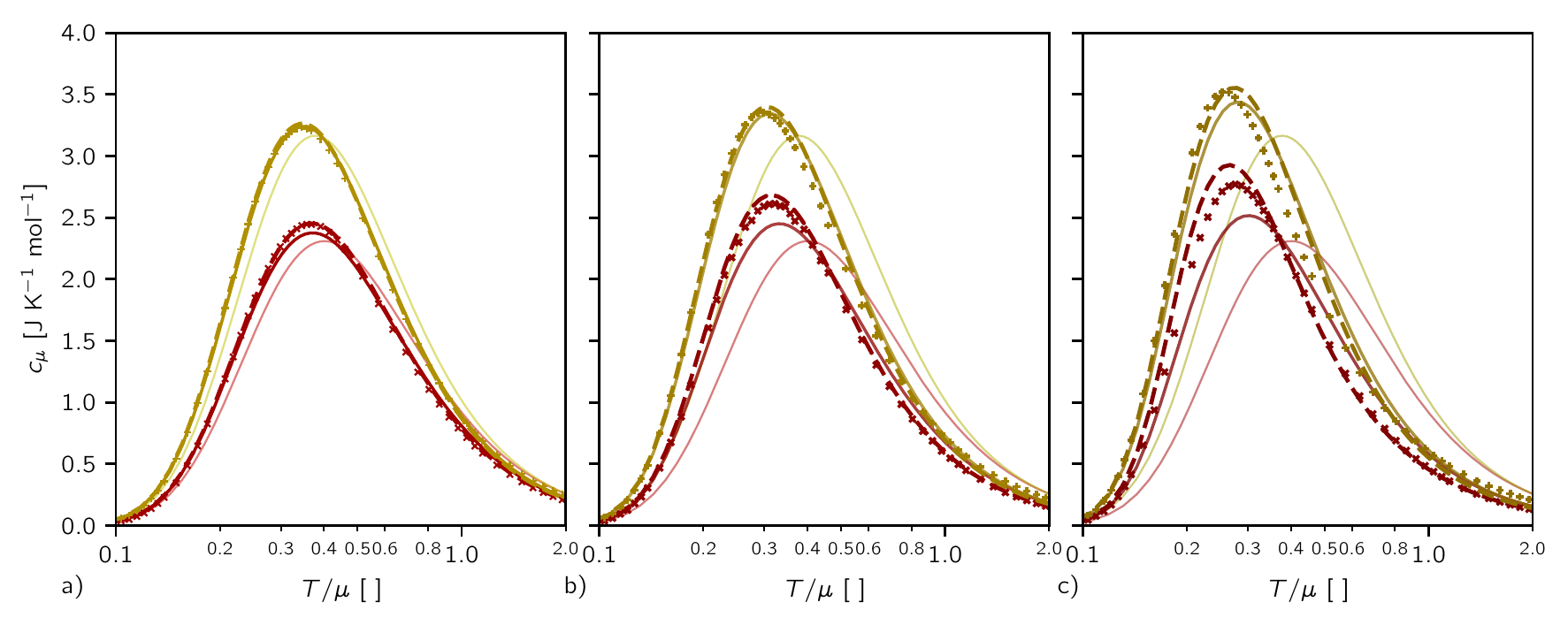}}
\caption[$\;\;$ Specific heat]{{\textbf{Specific heat}: simulation data {\it vs} $T/\mu$ for the electrolyte (yellow crosses) and primitive magnetolyte (red crosses) with corresponding Debye--H\"{u}ckel theory (yellow full line and red full line) and its pairing extension (yellow dashed line and red dashed line). The pairing correction is negligible for the weakly coupled system (left panel) but its significance increases as the Coulomb interaction strengthens. The exact specific heat for non-interacting particles is shown for reference (light yellow and light red lines. {\bf panel (a)} with $\zeta=0.27$ (see text), {\bf panel (b)} with $\zeta=0.54$, as for HTO, {\bf panel (c)} $\zeta=0.71$ as for DTO.}}
\label{fig:C_mu_simulations}
\end{figure*}

To illustrate the expectations based on Debye--H\"{u}ckel theory, we show in Fig.~\ref{fig:C_mu_simulations}, simulation and theory for interaction parameters $\zeta=0.27,0.54$ and $0.71$, the latter two corresponding to HTO and DTO and the first to a fictitious weakly interacting XTO, with half the pair-wise Coulomb energy of HTO. Data is shown for a primitive magnetolyte and an electrolyte in each case and is plotted as a function of $\frac{k_BT}{\mu}$. In this form, the evolution in the data is uniquely due to the changing interaction strength. Also shown in each figure as a reference  is the data for the non-interacting lattice gas. 

From the XTO results one can see that Debye--H\"{u}ckel theory does approach an exact description of the specific heat in the weakly interacting limit for the electrolyte. For the magnetolyte, although the theory is excellent, a small discrepancy between simulation and data can still be observed. This discrepancy between magnetolyte and electrolyte is because in the former charge pairs form a stronger correction (see Appendix \ref{AppPairing}). A smaller part of the discrepancy is because we have the additional approximation of including the vacuum entropy within the Pauling approximation\cite{Nagle1966a,Castelnovo2011}. 
The error of Pauling approximation is due to correlations on the level of loops of six spins and longer. 
The loops also cause an error of similar order between our approximate entropy in Eq.~(\ref{entropy})
and the full entropy of spin ice with monopoles, and their contribution to spin ice entropy was estimated in Ref.~[\onlinecite{Isakov2004b}].

As the interactions increase, the specific heat peak sharpens and moves to lower temperature.
Deviations between Debye--H\"{u}ckel theory and simulation develop as the theory correctly predicts the shift in peak position, but underestimates the sharpening. This sharpening is captured by our magnetolyte-specific pairing theory. However, for the interaction strengths of the real materials there remains excellent qualitative agreement which indeed becomes quantitative at both high and low temperature. In comparison, the non-interacting model appears in error at both high and low temperature and gives only a poor qualitative description of the Schottky peak. Closer examination at low temperature shows an asymptotic approach towards the simulation results below $0.5$~K, as the monopole density falls to zero~\cite{Kaiser2014}. At high temperature the data always disagree, illustrating the importance of screening in a Coulombic system even in this limit.

\subsection{Experiment}

The heat capacities of the spin ices \DTO~and \HTO\,  were measured between 0.35 and 300 K by a heat-relaxation method, using a Quantum Design Physical Properties Measurement System (PPMS), equipped with a $^3$He option. An addendum measurement was made to evaluate the background of Apiezon Grease N and this contribution was subtracted from the data. Three repetitions were taken for each measurement to improve statistics. 

The equilibrium heat capacity may be modelled as the sum of nuclear (hyperfine), electronic and lattice heat capacities. The electronic contribution, which interests us here, may be isolated by correcting the total specific heat  for the nuclear and lattice specific heats. For \HTO\, the hyperfine parameter $A = 0.30$ K is accurately known \cite{Bramwell2001PRL}, but the correction for the nuclear specific heat becomes very large at low temperature. We estimate that systematic errors arising from the subtraction of this contribution to be negligible above $T\approx 0.8$ K and very small (a few per cent) down to 0.4 K. For \DTO\, the nuclear contribution is smaller, but the nuclear spin relaxation rate is quite slow and comparable to experimental timescales at low temperature. In other work \footnote{L. Bovo, unpublished.} we derived a robust estimate of the electronic contribution by comparing short time and long time measurements with measurements on an isotopically enriched sample with no nuclear contribution. 
We have also estimated bounds on the variation of specific heat caused by slow equilibration of the electronic spin system. In this way we estimate that systematic errors arising from such sources are entirely negligible above $T \approx 0.8$ K and very small (again no more than a few per cent) down to 0.4 K.  As regards the lattice (phonon) contribution we found that a $T^3$-type correction is inadequate for an accurate measurement of the electronic specific heat. Note that such a correction has been used in the past for spin ice materials \cite{Hiroi2003,Klemke2011}; but if the object is estimating entropy, as has usually been the case, then the error incurred is small. By detailed comparison with the case of \TbTO \cite{Ruminy2016}, we have established the accuracy of a correction for the lattice contribution that involves comparing with the measured heat capacities of non-magnetic \YTO\, and \LuTO\,.  These are iso-structural to the spin ices but have different Debye constants. A simple temperature-scaling gives a collapse of the phonon heat capacities over an acceptable range of temperature. Analysis of the corrections showed that systematic errors in the estimated electronic specific heat of the spin ices become negligible at temperatures less than $T \approx 8$ K. 

Summarising these factors, in Fig.~\ref{intro-figure}, we display the estimated electronic specific heat in the range 0.4--10 K, but emphasise that systematic errors can only be completely excluded in the more restricted range 0.8--8 K, as discussed above.  It is evident from the Figures that the theory with monopole pairing included is very satisfactory in both cases. In general the description is slightly more accurate for \HTO\ than for \DTO\ as would be expected from the fact that \HTO, having the larger $|\mu|$, and hence a lower charge density, corresponds more accurately to the Debye--H\"uckel linear approximation. Further discussion of systematic errors in the comparison of theory and experiment is given in App.~\ref{AppErrors}. 

\section{Conclusions}

Our theoretical analysis of the magnetolyte provides an accurate, yet economical, description of specific heat of spin ice, a complex frustrated magnet. 
We have demonstrated that the monopole picture provides a framework for thermodynamics of spin ice going beyond existing techniques such as mean field theory, single tetrahedron or Bethe lattice calculations \cite{Jaubert2013}. Using Debye--H\"{u}ckel theory and its extensions we find a quantitive description of spin ice over a full range of temperatures, whereas the previous work has only approximately dealt with low temperatures\cite{Castelnovo2011}. This kind of development has so far proved beyond the capacity of the spin picture. Hence in this regard, the magnetolyte takes us a step beyond the dipolar spin ice model from which it is derived. 

Our description of the magnetolyte by means of the grand potential affords an efficient approach to charge correlations that emphasises the role of the strongly correlated monopole vacuum in spin ice. The price one pays for this step however, is to neglect the finite energy scale of the bandwidth of Pauling states. This has important consequences, particularly at low temperature where ordering \cite{Melko2004} and corrections to spin ice physics \cite{Pomaranski2013} cannot straightforwardly be accounted for. We note that it is, at any rate, very remarkable that it is possible to describe a spin system in terms of its emergent {\it low-energy} fractionalised degrees of freedom across the {\it full} temperature range. We are not aware of another instance, that is not otherwise exactly soluble anyway, where this is possible.

Improved experimental technique and data analysis are indispensable for the precise match between theory and the experimental specific heat data.
As spin ice physics takes place in the 1 K temperature range, the magnetic degrees of freedom separate easily from lattice vibrations, allowing for measurements over a particularly wide range of temperatures. For example, the data in Figs.~\ref{intro-figure} and \ref{fig:C_mu_simulations} is over a range, $0.1 \lesssim \frac{k_BT}{\mu} \lesssim 2$, covering both the high and low temperature regimes. To obtain this range we have presented new experimental data for specific heat of \HTO~and \DTO~with an improved analysis regarding the subtraction of non-magnetic effects.

At the same time, our work shows that spin ice models and materials provide a remarkable testing ground for Coulombic lattice fluids. 
Due to its solid state host, the magnetolyte is unique among them in several aspects: exact charge symmetry, absence of solvent, and precise control of chemical potential.
The charge symmetry of magnetic monopoles makes it the best realisation of the restricted primitive model, which allows for many simplifications of analytical calculations. 

The absence of solvent allows us to model electrostatic correlations exclusively without having to consider solvatation effects, 
such as the temperature dependence of interactions due to varying dielectric constant. 
\mbox{Neither} is there an effect of electrostatic interactions on the chemical potential of the solvent, 
because the number of ground state (empty) sites of the magnetolyte is fixed by the number of charges. 
This means that the osmotic coefficient\cite{Pitzer1973B} is always unity.
Finally, the absence of solvent allows for the twenty-fold variation in temperature in our experiments, 
which is larger than the ratio between evaporation and freezing temperatures of most common solvents.

Unlike many electrolyte systems, spin ice provides a Coulomb fluid in the grand canonical ensemble which is the natural setting to observe charge density fluctuations.
The material parameters of spin ice determine the chemical potential of the magnetolyte, as a fixed parameter independent of temperature. 
There is a broad choice of the value of the chemical potential, as well as the Coulomb coupling, which can be further tuned by chemical pressure \cite{Zhou2012},
while staying remarkably stable under hydrostatic pressure. 
The standard state, used in chemistry to define chemical potential, is thus determined robustly in spin ice.

Moreover, monopoles have no kinetic energy unlike dissolved ions, as all kinetic energy is electronic, quantised and fully contained in the magnetic terms of the Hamiltonian.
This does away with the need to consider the evolution of the thermal de Broglie wavelength and the kinetic energy with temperature, unlike in the lattice electrolytes of Ref.~[\onlinecite{Kobelev2002}], thus further anchoring the chemical potential.

It should finally be noted that our definition of chemical potential of the charges differs from the usual chemical one in the choice of the reference state, which conventionally would be that of an appropriate ideal gas, with activity coefficient defined by $n=\exp(\beta\mu)/\gamma_\mathrm{id}$. Our reference state is a non-interacting lattice gas with on-site exclusion and statistical weights fixed from the spin ice manifold. In effect, we include energetic terms in $\gamma$ as defined in Eq.~(\ref{gamma-def}), while treating the entropic contribution of the hard core exclusion between charges separately. The activity coefficient with respect to an ideal gas is related to our approach by the transformation
\begin{eqnarray}
\gamma_\mathrm{id} = \frac{3}{4}\gamma + \exp(\beta\tilde{\mu}) + \frac{\exp(4\beta\tilde{\mu})}{4\gamma^3} \;.
\end{eqnarray}
As a consequence, the limiting behaviour of the magnetolyte is that of a stochastic lattice gas, rather than an ideal gas.

We note that the correspondence between `autoionization' of spin ice and of water (both liquid and ice), in Eq.~(\ref{scheme}), means that the results obtained are relevant to these two very important electrochemical systems. 
For water and ice, the usual approach of electrochemistry is to use the Gibbs potential 
and exploit the conservation of chemical species during the dissociation.  
This `chemical' approach differs from ours in that the number of `water molecules' is fixed as in the canonical ensemble; 
hence the chemical potentials of all species (including water) vary with temperature. 
It is straightforward to show that the chemical approach, 
combined with assumption of the Pauling entropy for pure water, 
gives identical results to those generated here. 
This is not an entirely trivial observation: at first sight the Dirac string correlations, 
in either spin ice or water ice, render the chemical approach, 
which is based on the statistical independence of chemical species, questionable. 
However our analysis shows that the Pauling approximation restores the independence of the (effective) chemical species 
in Eq.~(\ref{scheme}), and allows the standard method to be applied over a range of temperatures. 
We speculate that this result 
helps justify the application of chemical thermodynamics to the auto-ionisation of liquid water, 
which like its solid form (water ice) is far from being a passive solvent for hydrogen ions \cite{Ludwig2001}.

To conclude, spin ice is a rare example of an experimentally accessible grand-canonical Coulomb fluid with varying interaction strengths in which one can confront Debye--H\"{u}ckel theory and test systematic improvements to it.
Furthermore, other electrolyte effects, such as their non-linear response, can be observed in spin ice, as authors of this paper have previously shown theoretically\cite{Kaiser2015} and experimentally\cite{Paulsen2016}.
In the future, phenomena that could be probed in this model material include confinement of electrolytes\cite{Bovo2014} and the role of quenched disorder and glassiness in long-range interacting systems\cite{Sen2015}.

\begin{acknowledgements}
It is a pleasure to thank L.D.C.~Jaubert, M.J.P.~Gingras, A.~Sen, C.~Castelnovo and S.~Sondhi for many fruitful discussions about spin ice and its models, as well as for collaboration on related work. This work was in part supported by the Deutsche Forschungsgemeinschaft via grant SFB 1143. L.~Bovo was supported by the Leverhulme Trust through the Early Career Fellowship program (ECF2014-284).
\end{acknowledgements}

\appendix

\section{Further discussion of systematic errors in the comparison of theory with experiment}\label{AppErrors}
In addition to the sources of systematic error described in the main text (i.e. inaccuracies in the Debye--H\"uckel linear approximation, and in the experimental correction for the nuclear and phonon contributions to the specific heat) there are several more subtle sources of systematic error that appear in our comparison of theory and experiment.

The dipolar spin ice parameters describing \DTO~and \HTO~were originally estimated by fitting experimental data for specific heat divided by temperature, $c/T$, to numerical simulations of the dipolar spin ice model (DSI)\cite{denHertog2000}. These parameters were later used to infer the parameters of the magnetolyte model\cite{Castelnovo2008}. We have finally used these magnetolyte parameters to calculate the  specific heat within extended Debye--H\"uckel theory, which is then compared with experiment. 

Like the DSI, the magnetolyte model has three parameters: $\{Q,a,\mu\}$ where DSI has $\{g,a,J\}$. Here $Q$ is the monopole charge, $a$ is the cubic lattice parameter, $\mu$ is the monopole chemical potential, $g$ is the rare earth g-factor, and $J$ an exchange coupling. 
Small systematic differences between theory and experiment appear in approximating the real materials to DSI, in the original choices of $g$ and $a$, and in approximating the DSI to the magnetolyte model. Of these only the values of $g$ and $a$ can be freed from systematic errors by more accurate measurements, but this is barely worthwhile given the fundamental systematic differences between DSI, the magnetolyte model and the experimental systems. These factors contribute systematic errors of order 1\% in the comparison of theory and experiment for the specific heat.

The magnetolyte reproduces the specific heat of both experiment and the DSI to high accuracy above around $0.4$ K. Below this temperature the models differ as the DSI orders \cite{Melko2001} due to the finite band width of Pauling states. Spin correlations are modified by this energy scale and extra parameters are required in the DSI to describe neutron scattering at low temperature (for example) \cite{Yavorskii2008}. The physics related to this energy scale is completely neglected in the magnetolyte but our results show that it does not affect the monopole thermodynamics over the temperature range $0.4-10$ K.

\section{Discriminating between double defects in electrolytes and magnetolytes}\label{AppTheories}
\begin{figure} [ht]
\centering{\includegraphics[scale=1.0]{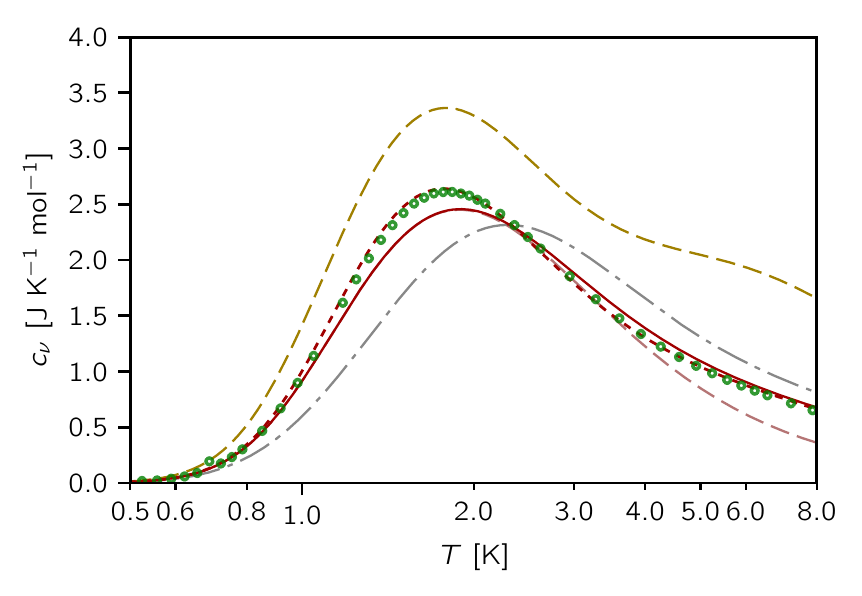}}
\caption[$\;\;$ Only DH theory and its extension can quantitatively explain specific heat of spin ice.]{{Comparing simulation specific heat of double defect dumbbell model (green circles) with a variety of theories illustrates different contributions to the specific heat: simulation data for $\zeta=0.54$ as for HTO, including double monopoles. Theoretical curves: Non interacting magnetolyte (gray dashed), Debye--H\"{u}ckel theory with single monopoles only (red dashed), Debye--H\"{u}ckel theory for the electrolyte including double charges (yellow dashed), Debye--H\"{u}ckel theory for magnetolyte including double charges (solid brown) and with pairing considered (dotted brown).
}}
\label{fig:C_theorydiff}
\end{figure}

In Fig.~\ref{fig:C_theorydiff} we show simulation data for the full magnetolyte, together with the corresponding Debye--H\"{u}ckel theory. 
Also shown are alternative theoretical approaches that capture the different many body effects at play: non-interacting theory,  Debye--H\"{u}ckel theory for the primitive magnetolyte and for an electrolyte including double charges. All fail to capture the simulation data as discussed in the main text. 
Including double charges for the electrolyte produces a clear second feature at higher temperature, corresponding to the thermal excitation of the second species. Although the effect of double monopoles is clearly observable in the simulation, such a pronounced double feature is not present in the magnetolyte as their weight is constrained by the vertex counting (see Eq.~\ref{entropy} and discussion). 

\section{Monte Carlo simulations}\label{AppSimulations}

To obtain the simulation data in this article, we performed Monte Carlo simulations of the dumbbell model of spin ice.
We used four types of Monte Carlo steps: single spin flips (\texttt{S}), monopole moves (\texttt{M}), charged worms (\texttt{C}), and loop flips (\texttt{L}). Single spin flips attempt to flip a random spin in the system ($2N_0$ times per step), which moves a charge or creates/destroys a nearest-neighbour $(+-)$ charge pair. We also keep a list of monopoles that we randomly choose from to propose a move to one of the neighbouring sites ($N_0$ times per step). Finally, the worm steps construct either a string of spins that flips while moving a charge across the system or a loop of spins that flips without changing. For our simulations, we used the order \texttt{SMSLMSMLSMSC} of MC steps for each sweep. 
We used Metropolis update scheme for all MC steps. 
The Coulomb energy was evaluated by Ewald summation with metallic boundary conditions at infinity. 
The specific heat was calculated using the fluctuation-dissipation theorem.
All the above-mentioned methods are further detailed in Refs.~[\onlinecite{Kaiser2014,Jaubert2011}]. 

The simulated system contains $L^3$ pyrochlore lattice unit cells, i.e. $8L^3$ charge sites and $16L^3$ spins. We adapted system size $L$ to be larger than twice the Debye screening length (see Table \ref{tableApp}). The system size increases fast with the lowering temperature as the monopole number density increases exponentially. As the memory cost of the simulation increases, we reduce the total number of steps taken.
Nevertheless, the worm algorithm ensures that the configuration space is sampled efficiently independent of the temperature and monopole density.

\begin{table}[htbp]\begin{center}
\begin{tabular}{rrrr}
$T$ & $\Delta T$ & $L$ &  \# MC sweeps \\ \hline
2.25--10 & 0.25 & 6 & 100000 \\
1.6--2.0 & 0.1 & 8 & 100000 \\
0.90--1.55 & 0.05 & 8 & 100000 \\
0.60--0.85 & 0.05 & 10 & 100000 \\
0.55--0.575 & 0.025 & 12 & 100000 \\
0.50--0.525 & 0.025 & 16 & 100000 \\
0.475 & 0.025 & 20 & 100000 \\
0.450 & 0.025 & 24 & 10000 \\
0.425 & 0.025 & 32 & 10000 \\
0.400 & 0.025 & 40 & 10000
\end{tabular}
\caption{Parameters of our MC simulations for \DTO: temperature range, temperature step, system size, and the total number of sweeps.}
\label{tableApp}
\end{center}\end{table}

\begin{widetext}
\section{Pairing theory}\label{AppPairing}

This appendix serves to describe methods how to include pairing as a next order correction in electrolytes and magnetolytes.
I show that unlike the Debye--H\"{u}ckel theory, the pairing theory is not transferable between electrolytes and magnetolytes due to their different short range structure.

\subsection{Electrolytes}
Pairing in lattice electrolytes has previously been described in Ref.~[\onlinecite{Kobelev2002}], 
which used the Bethe approximation for the monomer-dimer model as given by Nagle\cite{Nagle1966b}. We adapt this approximation to include orientable dimers, i.e. dipoles.
The number of configurations available for positive and negative charges and for their oriented nearest-neighbour dipoles on a lattice is
\begin{align}
W &= \left[ \binom{N}{\frac{Nn_1}{2}\quad \frac{Nn_1}{2} \quad Nn_b\quad N(1 - n_1  - n_b)} (2q)^{N n_b} \right] \\
&\times
\left[ \frac{n_b}{2 q} \left( 1 - \frac{n_b}{q} \right)^{q-1} \right]^\frac{N n_b}{2} 
\left[ \left( 1 - \frac{n_b}{q} \right)^q \right]^\frac{N(1-n_b)}{2} \;,
\end{align}
where the first bracket describes the possible placings of positive charges, negative charges, and their bound states; 
$q=4$ is the connectivity of the lattice; the latter brackets describe the compatibility of dimers and monomers with their neighbouring sites.

The corresponding entropy reads
\begin{align}
\frac{S}{N k_B} = \frac{1}{N}\log(W) 
&= -n_1 \log \frac{n_1}{2} - (1-n_1-n_2-n_b) \log (1-n_1-n_b) \\
&\quad - \frac{n_b}{2} \log\frac{n_b}{8} + \left(2 - \frac{n_b}{2}\right)  \log\left(1-\frac{n_b}{4}\right)
\end{align}
and free energy
\begin{align}
\frac{\Omega}{N} &=  \frac{U}{N} - \sum_i n_i \mu_i - \frac{TS}{N} \\
&= -n_1 \tilde{\mu}  - \frac{n_b}{2} \left(2\mu + kT\log K_E\right)
 - k_B T \Bigg[ -n_1 \log \frac{n_1}{2} - (1-n_1-n_2-n_b) \log (1-n_1-n_b) \\
&\qquad - \frac{n_b}{2} \log\frac{n_b}{8} + \left(2 - \frac{n_b}{2}\right)  \log\left(1-\frac{n_b}{4}\right)\Bigg]\;,
\end{align}
where the chemical potential of the pairs follows from their chemical equilibrium with the free charges set by the truncated Ebeling association constant $K_E=\exp(-\beta U) + \exp(\beta U) - 2 - \frac{(\beta U)^2}{2}$.
The continuous Ebeling constant is the integral of the previous expression over the whole space, which exactly captures the excess correlations in electrolytes to order $l_D^2$, while preserving the previously derived DH theory\cite{Ebeling2002}. While it formally treats only the $+-$ association, Ebeling's theory in fact includes all correlations of this order, even between like charges. 
Due to the effort needed to calculate entropy of pairs of all sizes allowed by the diamond lattice, we only consider nearest neighbour pairs and truncate $K_E$.

The free energy can be minimized with respect to the number densities, yielding
\begin{align}
n_1 &= \frac{6 e^{\beta\tilde\mu} }{ 1 + 2 e^{\beta\tilde\mu} + 2 \sqrt{ (1 + 2 e^{\beta\tilde\mu} )^2 + 6 K_E e^{2\beta\mu} } } \\
n_b &= \frac{ (1 + 2 e^{\beta\tilde\mu})^2 + 4 K_E e^{2\beta\mu} - (1 + 2 e^{\beta\tilde\mu}) \sqrt{(1 + 2 e^{\beta\tilde\mu})^2 +  6 K_E e^{2\beta\mu} } }{ \frac{1}{2}(1 + 2 e^{\beta\tilde\mu})^2 +  4 K_E e^{2\beta\mu} }
\end{align}
which limit to the DH theory for $K_E\rightarrow0$.
This approach can be easily extended to include double charges.
The specific heat is obtained from Eq.~(\ref{cmu_calculation}) as in the main text.

\begin{figure}[htb]
\begin{center}
\includegraphics[width=\textwidth]{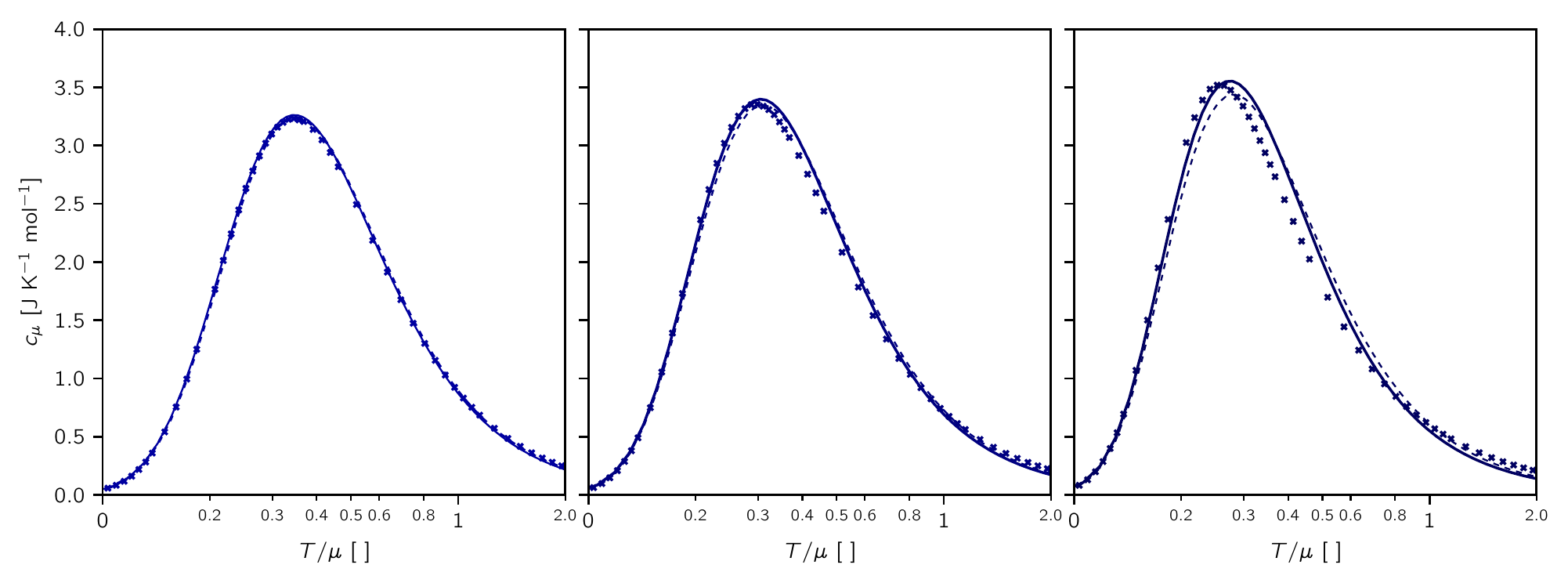} 
\caption{Pairing theory for lattice electrolytes. Specific heat curves for electrolytes with $\mu$ and $\zeta$ of XTO, HTO, and DTO.
Points are simulation results, dashed line the DH theory, full line the pairing theory. Temperature is given in units of the chemical potential.}
\label{fig:el_pairing}
\end{center}
\end{figure}

\begin{figure}[htb]
\begin{center}
\includegraphics[width=\textwidth]{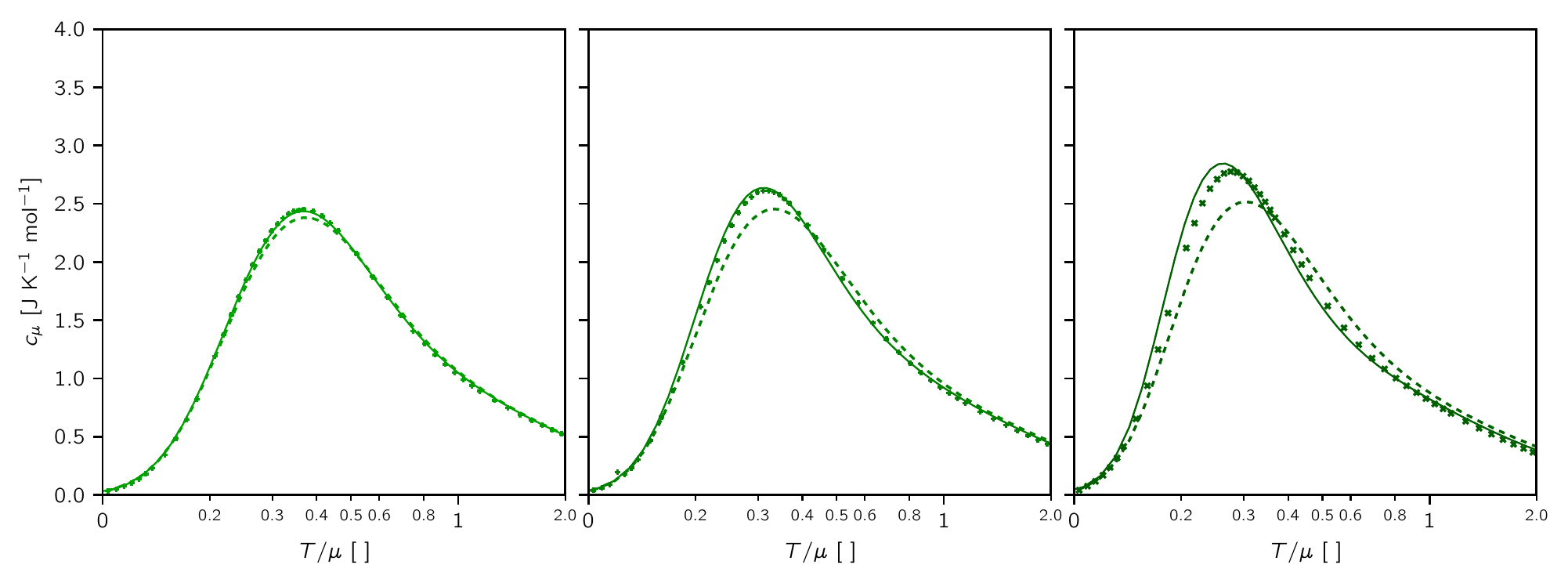} 
\caption{Pairing theory for the double-defect magnetolyte. Specific heat curves for $\mu$ and $\zeta$ corresponding to XTO, HTO, and DTO.
Points are simulation results, dashed line the DH theory, full line the pairing theory. Temperature is given in units of the chemical potential.}
\label{fig:si_dd_pairing}
\end{center}
\end{figure}

\subsection{Magnetolytes}

For spin ice, a different approach has to be taken, because every charge state is underpinned by multiple spin configurations. 
Nagle's argument about monomer--dimer compatibility fails completely because the compatibility of every vertex with its neighbour is fully determined by the orientation of the spin connecting them,
which is always compatible with half of the vertex states.
This makes compatibility of pairs with neighbouring charges easier to achieve which, in turn, promotes pairing in comparison with electrolytes.
As an alternative, the calculation can be performed within the scope of a two site (7 spins, with $2^7=128$ configurations) approximation.
Six of the spins are shared with neighbouring tetrahedra, while one spin is internal.
The following number of spin configurations corresponds to given charge configurations.\\[1em]

\begin{small}
\begin{tabular}{ c c c c c c c c c }
  Tetrahedra & \bn\bn & \bn\bbp & \bn\bbm & \bbm\bbp & \bbp\bbp & \bbm\bbm  \\
  No. states & 18 & 24 & 24 & 20 &  6 & 6  \\
  Eff. chem. pot. & 0 & $\mu$ & $\mu$ & $2\mu+kT\log(K_{+-})$ & $2\mu+kT\log(K_{++})$ & $2\mu+kT\log(K_{--})$ \\[2em]

  \bn\bpp & \bn\bmm & \bbm\bpp &  \bbp\bmm & \bbp\bpp & \bbm\bmm & \bpp\bmm \\
  6 & 6 & 6 & 6 & 2 & 2 & 2 \\
   $4\mu$ & $4\mu$ & $5\mu+kT\log(K_{-\overset{+}{+}})$ & $5\mu+kT\log(K_{+\overset{-}{-}})$ & $5\mu+kT\log(K_{+\overset{+}{+}})$ & $5\mu+kT\log(K_{-\overset{-}{-}})$ & $8\mu+kT\log(K_{\overset{+}{+}\overset{-}{-}})$ 
\end{tabular}\\[1em]
\end{small}

The electrostatic interactions can once again be included by replacing the chemical potential $\mu$ with the effective chemical potential $\tilde\mu$ from the DH theory.
This constrains the association constants to limit to unity ($K\rightarrow1$) in order to recover DH theory in the single-vertex non-pairing case, which excludes the Ebeling approach from above, and therefore we adopt Bjerrum-like association constants below.
Other constraints on the association constants are the charge symmetry $K_{--}=K_{++}$ and the quadratic scaling with ionic strength $K_{\overset{+}{+}\overset{-}{-}}=K_{+\overset{-}{-}}^2=K_{+-}^4$.

The total number of configurations reads 
\begin{align}
W &= \left( \frac{1}{2} \right)^{3N_0} \frac{N_0 !}{N_1!N_2!\dots N_{128}!} \;,
\end{align}
where the 128 vertices are assigned the following charge identities
$N_1=N_2=\dots N_{18}=N_{\varnothing\varnothing}/18$, 
$N_{19}=\dots N_{42}=N_{\varnothing+}/24$,
$N_{43}=\dots N_{66}=N_{\varnothing-}/24$,
$N_{67}=\dots N_{86}=N_{+-}/20$,
$N_{87}=\dots N_{92}=N_{++}/6$,
$N_{93}=\dots N_{98}=N_{--}/6$,
$N_{99}=\dots N_{104}=N_{\varnothing\overset{+}{+}}/6$,
$N_{105}=\dots N_{110}=N_{\varnothing\overset{-}{-}}/6$,
$N_{111}=\dots N_{116}=N_{-\overset{+}{+}}/6$,
$N_{117}=\dots N_{122}=N_{+\overset{-}{-}}/6$,
$N_{123}= N_{124}=N_{+\overset{+}{+}}/2$,
$N_{125}= N_{126}=N_{-\overset{-}{-}}/2$,
$N_{127}= N_{128}=N_{\overset{-}{-}\overset{+}{+}}/2$, 
where the first and the second indices describe the charge at the respective diamond lattice sites.

This leads to the entropy per tetrahedron
\begin{align}
\frac{S}{(N/2) k_B} = \frac{1}{N/2}\log(W) 
&= -n_{\varnothing\varnothing} \log\left(\frac{8n_{\varnothing\varnothing}}{18}\right) - n_{\varnothing+} \log\left(\frac{8n_{\varnothing+}}{24}\right) - n_{\varnothing-} \log\left(\frac{8n_{\varnothing-}}{24}\right) \\
&\quad -n_{+-} \log\left(\frac{8n_{+-}}{20}\right) - n_{\varnothing\overset{+}{+}} \log\left(\frac{8n_{\varnothing\overset{+}{+}}}{6}\right) - n_{\varnothing\overset{-}{-}} \log\left(\frac{8n_{\varnothing\overset{-}{-}}}{6}\right) \\ 
&\quad- n_{++} \log\left(\frac{8n_{++}}{6}\right) - n_{--} \log\left(\frac{8n_{--}}{6}\right) - n_{-\overset{+}{+}} \log\left(\frac{8n_{-\overset{+}{+}}}{6}\right) \\
&\quad- n_{+\overset{-}{-}} \log\left(\frac{8n_{+\overset{-}{-}}}{6}\right)  - n_{+\overset{+}{+}} \log\left(\frac{8n_{+\overset{+}{+}}}{2}\right) - n_{-\overset{-}{-}} \log\left(\frac{8n_{-\overset{-}{-}}}{2}\right) \\
&\quad- n_{\overset{+}{+}\overset{-}{-}} \log\left(\frac{8n_{\overset{+}{+}\overset{-}{-}}}{2}\right)
\end{align}

If the chemical potentials for pairs of tetrahedra were simply sums of their components' chemical potentials, the free energy would factorise to the previously used single-vertex form. 
This factorisation follows from the fact that we have not introduced any additional correlations 
(as we do not include any loop which would have a minimal size of 6 tetrahedra).
This imposes the following relations on the single vertex densities
\begin{align}
n_{\varnothing\varnothing}&=n_{\varnothing}^2 \;,\; n_{\varnothing+}=n_{\varnothing-}=2n_{\varnothing}n_+ \;,\; n_{+-}=\frac{5}{2}n_+n_- \;,\; n_{++}=n_{--}=\frac{3}{4}n_+^2 \;,\;  \\
n_{\varnothing\overset{+}{+}}&=n_{\varnothing\overset{-}{-}}=2n_{\varnothing}n_{\overset{+}{+}}  \;,\;  n_{+\overset{-}{-}}=n_{-\overset{+}{+}}=3 n_{+}n_{\overset{-}{-}} \;,\; n_{+\overset{+}{+}}=n_{-\overset{-}{-}}=n_{+}n_{\overset{+}{+}} \;,\; n_{\overset{+}{+}\overset{-}{-}} = 4n_{\overset{+}{+}} n_{\overset{-}{-}}
\end{align}
and the symmetric relations are due to the macroscopic electroneutrality $n_+=n_-$,  $n_{\overset{+}{+}}=n_{\overset{-}{-}}$.

We are now faced with the choice of the association constant. The simplest choice is to use the Boltzmann weight of the nearest neighbour charge pair $K_{+-}=\exp(-\beta U_{NN})=\exp(2\ell_T/a)\overset{\mathrm{def.}}{=}K$, $K_{++}=K_{--}=\exp(\beta U_{NN})=\exp(-2\ell_T/a)=1/K$, $K_{-\overset{+}{+}} = K_{+\overset{-}{-}} = K^2$, $K_{+\overset{+}{+}} = K_{-\overset{-}{-}}=1/K^{2}$, $K_{\overset{+}{+}\overset{-}{-}}=K^4$. We also assume that all charges keep their DH correction to the chemical potential, yielding the free energy
\begin{align}
\frac{\Omega}{N/2} &=  \frac{U}{N/2} - \sum_i n_i \mu_i - \frac{TS}{N/2} \\
&= -\mu n_{\varnothing+} - (2\mu+kT\log K) n_{+-} - (2\mu-kT\log K) n_{++} \\
&\quad- 4 \mu n_{\varnothing\overset{+}{+}} - (5\mu+2 kT\log K) n_{-\overset{+}{+}} - (5\mu-2 kT\log K) n_{+\overset{+}{+}} - (8\mu+4 kT\log K) n_{\overset{+}{+}\overset{-}{-}}  \\
&\quad- k_B T \Bigg[ n_{\varnothing\varnothing} \log\left(\frac{8n_{\varnothing\varnothing}}{18}\right) 
+ 2n_{\varnothing+} \log\left(\frac{8n_{\varnothing+}}{24}\right)
+ n_{+-} \log\left(\frac{8n_{+-}}{20}\right) 
+ 2n_{++} \log\left(\frac{8n_{++}}{6}\right) \\ 
&\qquad + 2n_{\varnothing\overset{+}{+}} \log\left(\frac{8n_{\varnothing\overset{+}{+}}}{6}\right)
+ 2n_{-\overset{+}{+}} \log\left(\frac{8n_{-\overset{+}{+}}}{6}\right) 
+ 2n_{+\overset{+}{+}} \log\left(\frac{8n_{+\overset{+}{+}}}{2}\right) 
+ n_{\overset{+}{+}\overset{-}{-}} \log\left(\frac{8n_{\overset{+}{+}\overset{-}{-}}}{2}\right) \Bigg] \;.
\end{align}
This free energy can be truncated accordingly to include only singly charged configurations.

We minimize again with respect to the densities
\begin{align}
n_{\varnothing\varnothing} &= 9/Z \qquad
n_{\varnothing+} = n_{\varnothing-} = 12 e^{\beta\tilde\mu}/Z \qquad
n_{+-} = 10 K e^{2\beta\tilde\mu}/Z \qquad
n_{++} = n_{--} = 3 K^{-1} e^{2\beta\tilde\mu}/Z \\
n_{\varnothing\overset{+}{+}} &= n_{\varnothing\overset{-}{-}} = 3 e^{4\beta\tilde\mu}/Z \qquad\qquad\quad
n_{-\overset{+}{+}} = n_{+\overset{-}{-}} = 3 K^2 e^{5\beta\tilde\mu}/Z \\
n_{+\overset{+}{+}} &= n_{-\overset{-}{-}} = K^{-2} e^{5\beta\tilde\mu}/Z \qquad\qquad
n_{\overset{+}{+}\overset{-}{-}} = K^{4}e^{8\tilde\beta\mu}/Z \\
&\text{where } Z = 9 + 24 e^{\beta\tilde\mu} + 10 K e^{2\beta\tilde\mu} + 6 K^{-1} e^{2\beta\tilde\mu} + 6 e^{4\beta\tilde\mu} + 6 K^2 e^{5\beta\tilde\mu} + 2 K^{-2} e^{5\beta\tilde\mu} + K^4 e^{8\beta\tilde\mu} \;,
\end{align}
which translates to free charge densities
\begin{align}
n_{\varnothing} &= \sqrt{n_{\varnothing\varnothing}} = \frac{3}{\sqrt{Z} } \qquad
n_1 = 2\frac{n_{\varnothing+}}{2n_{\varnothing}} = \frac{4 e^{\beta\tilde\mu}}{\sqrt{ Z } } \qquad
n_2 = 2\frac{n_{\varnothing\overset{+}{+}}}{2n_{\varnothing}} = \frac{e^{4\beta\tilde\mu}}{\sqrt{ Z} } \;.
\end{align}
The specific heat is again obtained using the procedure outlined in Eq.~(\ref{cmu_calculation}) of the main text.

The limitation of this approach is that the we assume that even charges in dipoles keep the DH form of screening, which partially double-counts the electrostatic interactions. 
A fully consistent approach would require a study of mean-field screening of all the nearest-neighbour charge configurations appearing in our expansion, as outlined for electrolytes in Ref.~[\onlinecite{Fisher1993}].

For both electrolytes and magnetolytes, pairing improves on the specific heat description using DH theory.
However, the description of pairing in the two scenarios differes significantly, 
which demonstrates that short-range structure and the emergent nature of spin ice differs from lattice electrolytes,
while long-range properties of spin ice and electrolytes match well.
\end{widetext}

\end{document}